\documentclass[aps,prb,twocolumn,groupedaddress,reprint]{revtex4}
\usepackage{graphicx}  
\usepackage{dcolumn}   
\usepackage{bm}        
\usepackage{amssymb}   
\usepackage{xfrac}

\usepackage{color}
\definecolor{myBlue}{rgb}{0,0,1.0}
\definecolor{myGreen}{rgb}{0,0.8,0}

\begin{document}



\title{Bimodal Magnetic Force Microscopy with Capacitive Tip-Sample Distance Control}

\author{J.~Schwenk$^1$}
\author{X.~Zhao$^1$}
\author{M.~Bacani$^1$}
\author{M.~Marioni$^1$}
\author{S.~Romer$^1$}
\author{H.~J.~Hug$^{1,2}$}
\affiliation{$^1$Empa, Swiss Federal Laboratories for Materials Science and Technology, CH-8600 D\"{u}bendorf, Switzerland.}
\email[Corresp. author. E-mail:~]{johannes.schwenk@empa.ch}
\affiliation{$^2$Department of Physics, University of Basel, CH-4056 Basel, Switzerland}


\begin{abstract}
A single-passage, bimodal magnetic force microscopy technique optimized for scanning samples with arbitrary topography  is discussed. 
A double phase-locked loop (PLL) system is used to mechanically excite a high quality factor cantilever under vacuum conditions on its first mode and via an oscillatory tip-sample potential on its second mode. 
The obtained second mode oscillation amplitude is then used as a proxy for the tip-sample distance, and for the control thereof. 
With appropriate $z$-feedback parameters two data sets reflecting the magnetic tip-sample interaction and the sample topography are simultaneously obtained. 

\end{abstract}

\pacs{}

\maketitle

%
Magnetic Force Microscopy is a versatile technique to image local magnetic fields with high spatial resolution \cite{SPM}. 
It is achieved by scanning an ultrasharp, high-aspect ratio magnetic tip along the surface of the sample at small tip-sample distances under vacuum conditions. 
The latter is required for a high Q-factor of the cantilever, which in turn allows obtaining high sensitivity to small magnetic forces \cite{Schwenk:2014fe}. 
Usually a dual passage method is used, where each measurement line is scanned twice \cite{Hosaka:1992,Giles:1993fu}. 
A first scan is carried out in the intermittent contact mode and reveals the topography. 
A subsequent scan takes place without tip-sample contact at a user selected lift height, following the topography data recorded in the first scan.
However, the use of the intermittent contact mode in vacuum remains challenging\cite{Schwenk:2014fe}. 

Recently single passage measurement methods have been reported that use bimodal cantilever excitation suitable for operation in air \cite{Li:2009ff} and in vacuum \cite{Schwenk:2014fe}. 
They rely on the ability to separate magnetic from non-magnetic (van der Waals or electrostatic) forces on the basis of their different decay lengths. 
But magnetic fields of small magnetic structures can have the same decay length as van der Waals forces, making the separation of magnetic and topography-induced forces difficult in these situations. 
Moreover, scanning at constant average height, as is often convenient for quantitative data analysis \cite{Schendel00, Schmid:2010fz}, or operation at larger tip-sample distances becomes challenging, because the situation arises where the (longer-ranged) magnetic forces dominate the (shorter-ranged) topographical forces.
The latter can then no longer be used for tip-sample distance control.
Additional problems arise if measurements are performed at different temperatures or external magnetic fields.
Both change the resonance frequency of the free cantilever, requiring a re-adjustment of the frequency shift set-point used for recording the topography. 
Although such reset is possible, it is often impractical, e.g.~when the magnetization of the magnetic coating on the cantilever beam settles slowly over time leading to a corresponding creep of the free resonance frequency. 

\begin{figure}
\includegraphics{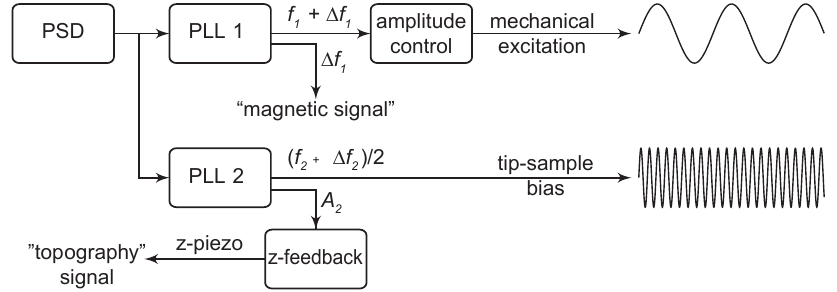}
\caption{\small schematic of the dual-PLL system required for bimodal oscillation of high quality factor cantilevers. 
The first PLL mechanically drives the cantilever on its first mode, and tracks shifts of its resonance frequency. 
The second PLL excites the cantilever via an oscillatory electric field at half the resonance frequency of its second mode. 
The $z$-feedback then keeps the obtained second mode oscillation amplitude constant to map the sample topography. 
 The required $z$-travel then reflects the topography of the sample. } 
\label{Fig:setup}
\end{figure}

Here we propose a  single passage measurement technique that overcomes these limitations. 
Figure\,\ref{Fig:setup} depicts its schematic setup \cite{ZHI}. 
As in our previous work \cite{Schwenk:2014fe} a first phase-locked loop (PLL) is used to drive mechanically the cantilever on its first flexural resonance $f_1$, typically with an oscillation amplitude $A_1=10$\,nm (zero-to-peak), chosen to optimize the ratio between the measured magnetic force induced frequency shift and the frequency noise caused by thermal fluctuations \cite{Rugar:2001hm, Schwenk:2014fe}. 
%
%
Unlike our previous work \cite{Schwenk:2014fe} the second cantilever oscillation mode at $f_2=6.27 \cdot f_1$ \cite{Butt:1995ui} is {\em not} driven mechanically but by applying an oscillatory tip-sample bias. 
In general, a bias of the form  $U(t) = U_{dc} + U_{ac}\cos (2\pi f_{ac}t)$  generates an electrostatic force given by 
\begin{eqnarray}
F_E(z,t) &=& \frac{1}{2}\frac{\partial C(z)}{\partial z}  \cdot \left[ \right. U_{dc}^2 + 2 U_{dc} U_{ac} \cos (2\pi f_{ac}t) \nonumber \\
           &+&  \left. U_{ac}^2 \cos ^2 (2\pi f_{ac}t) \right],
\label{eq:Eforce}
\end{eqnarray}
where $C(z)$ is the distance dependent tip-sample capacity, $U_{dc} = U_{dc}^{(K)}+U_{dc}^{(a)}$ is the sum of the contact and applied potential, and $U_{ac}$ is the amplitude of the potential modulation.
We see from Eq.~(\ref{eq:Eforce}) that $F_E$ has components at frequency $f_{ac}$ and $2f_{ac}$, the latter being:
\begin{equation}
F_{E,2f_{ac}}(z) = \frac{1}{4}\frac{\partial C(z)}{\partial z}  \cdot U_{ac}^2 \, .
\label{eq:F_E2}
\end{equation}
In particular,  a cantilever oscillation will be induced at $2f_{ac}$ that is proportional to $\partial C/\partial z$ but independent of $U_{dc}$  (and thus also insensitive to contact potentials).
That is significant because $\partial C/\partial z$ carries information of the tip-sample distance, so that the amplitude at $2f_{ac}$ can be a measure thereof. 
A spatial dependence of the contact potential $U_{dc}^{(K)}$ would lead to a corresponding variation of the first (and second) mode frequency shift, unless the Kelvin potential is compensated through a suitable implementation of a Kelvin feedback loop.   

By setting $f_{ac}=\sfrac{1}{2}\cdot f_2$, i.e.~half the second mode resonance frequency, resonance amplification  ensures a conveniently large amplitude $A_2$, but a second PLL  is needed to track $2f_{ac}$ as shown in Fig.\,\ref{Fig:setup}.
The latter requirement arises because magnetic forces acting on the tip and changes of the tip-sample distances, generate frequency shifts that can easily be larger than the width of the resonance peak and thus would significantly change the force-to-amplitude transfer function. 

\begin{figure}
\includegraphics{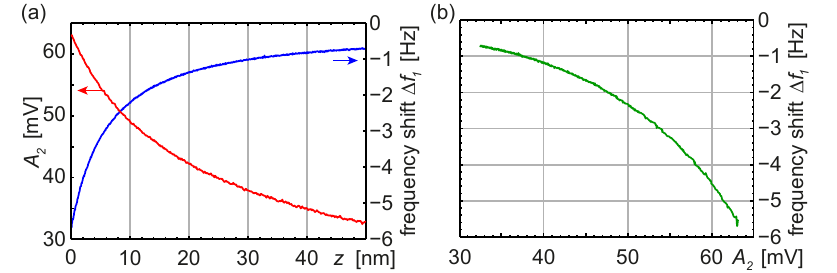}
\caption{\small (a) Second mode amplitude $A_2(z)$ (red line), and first mode frequency shift $\Delta f_1(z)$ (blue line) as a function of tip sample distance.
(b) The measured $\Delta f_1$-signal is re-plotted as a function of $A_2$. } 
\label{Fig:A2-df1}
\end{figure}

With the above setup we can obtain tip-sample distance dependent  $A_2(z)$ curves, a representative example of which is given in Fig.~\ref{Fig:A2-df1}(a), red line.
Note that  the first mode resonance frequency $\Delta f_1(z)$  can be measured simultaneously with the measurement of $A_2(z)$ -- cf.~Fig.~\ref{Fig:A2-df1}(a), blue line.
The monotonicity of $A_2(z)$ shows that it is suitable for controlling the tip-sample distance, i.e.~modifying the value of $z$ by $\Delta z$ until $A_2(z)$ equals a predefined setpoint, provided the dielectric response of the sample remains constant within the scan area. Further, the quality factor $Q_2$ must remain constant. The latter could change if for example magnetic dissipation occurred. That would however also affect the first mode quality factor $Q_1$, which is not the case for our measurements.   

Conveniently the slope of $A_2(z)$ increases with decreasing $z$, whereby the signal-to-noise ratio of the measured $A_2(z)$ is improved. 
This facilitates a faster control of the tip-sample distance when it is of the same order or smaller than the height variation of topographical features.
Measuring with a $z$-feedback that rapidly adapts to the local conditions implies that $A_2(z)$ is (in an ideal case) constant, and that the map of the corresponding $\Delta z$ is a measure of the topography ({\em constant local height} imaging).
Conversely, if the $z$-feedback is disabled, $A_2(z)$ should a priori vary with position in accordance with the topography. Note however, that by the unavoidable drift of the tip-sample distance the latter will change not only locally, but on average. Such drift, but not local variations of the tip-sample distance, will be corrected if the $A_2(z)$-based control is retained but made sufficiently slow. With a slow $z$-feedback mode a type of {\em constant average height} imaging mode is obtained, wherein $A_2(z)$ is a measure of topography. 
Importantly, very small to almost arbitrarily large average distances from the sample surface can be maintained, on account of the large range where $\frac{\partial C(z)}{\partial z}$ varies. 
This represents a major advantage to the aforementioned bimodal technique \cite{Schwenk:2014fe}. 
%

\begin{figure}
\includegraphics{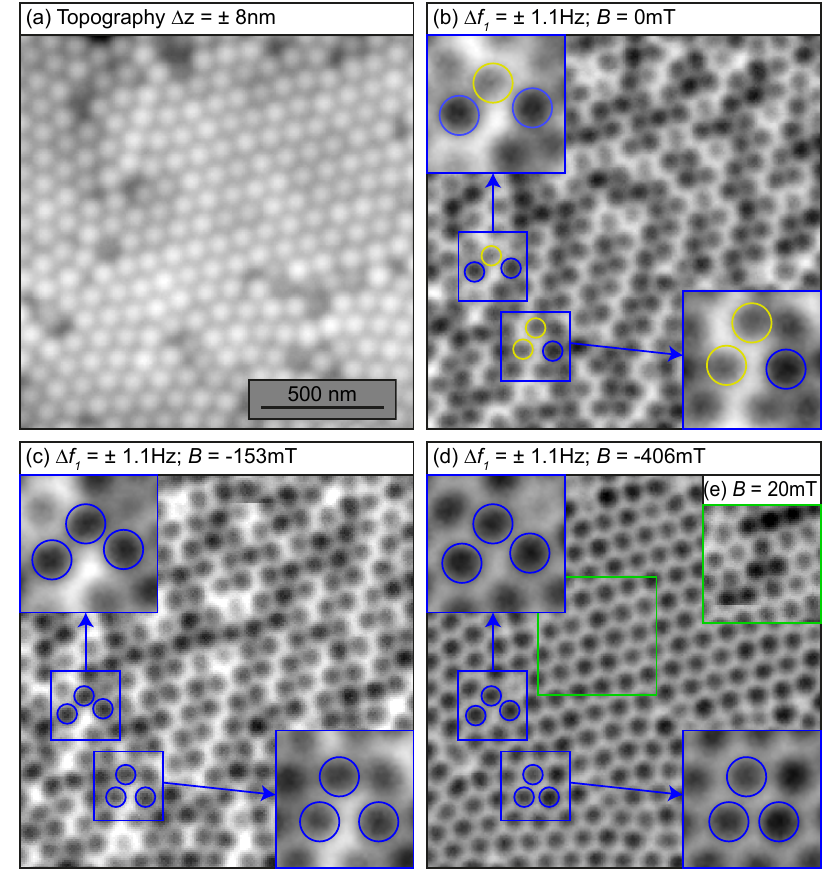}
\caption{\small 
	Data obtained with $A_2$ kept constant, i.e. using a fast $z$-feedback. 
	(a) Signal from the topography of the sample, i.e. the $\pm8\,$nm $z$-travel required to keep $A_2$ constant during scanning. 
	(b) $\Delta f_1(x,y)$ MFM data recorded simultaneously with (a) in zero field. The yellow/blue circles highlight dots with an up/down magnetization. 
	(c) MFM image taken at -153\,mT. 
	An area between the dots with an up magnetization is visible between the three domes inside the blue frames.
	(d) MFM data acquired in a field of -406\,mT that saturates the magnetic layer.
	(e) MFM measurement at +20\,mT of the area framed in green in (d).} 
\label{Fig:MFM_const_A2}
\end{figure}

%
To investigate the characteristics of this control mode in greater practical detail we work with a  Ta(5\,nm)/Pt(5\,nm)/[Co(0.4\,nm)/Pt(0.7\,nm)]$_4$/Pt(3.5\,nm)  multilayer deposited onto a hexagonal array of domes in an anodic alumina template with 100\,nm period, similar to Ref.~\cite{Piraux:2012}. 
This sample provides small magnetic features, but also large topographical ones which typically constitute a major difficulty for high-resolution magnetic force microscopy. In prior work \cite{Hauet:2014fc} the topography of such a sample was measured in air, with the intermittent contact mode. The typical bump-to-bump height variation was found to be around 5\,nm, but at defects much larger height variations of up to $\pm 8$\,nm  -- cf. Fig.\,\ref{Fig:MFM_const_A2}(a) -- were measured. 

A Team-Nanotec cantilever without coating, with a length of $225\,\mu$m, a width of $35\,\mu$m, and a nominal stiffness of $0.7\,$N/m was used.  Its high aspect ratio tip was made sensitive to magnetic stray fields by sputter coating one tip side with 2\,nm Ti and 4\,nm Co, and subsequently magnetizing it to have a north pole at the tip end.
The contact potential of 592\,mV was compensated (i.e.~$U_{dc}=0$ ). 
An oscillatory potential $U_{ac} = 500\,$mV was applied resulting in the $A_2(z)$ curve depicted in red in Fig.\,\ref{Fig:A2-df1}(a) for a range of $z$.

%
Figure \ref{Fig:MFM_const_A2}(a) and (b) show two simultaneously recorded channels of a first scan, taken in zero field, in which $A_2$  was kept constant with a fast feedback loop that varied $z$. It is an example of {\em constant local height} imaging.
The left panel is the map of $\pm8\,$nm $\Delta z$-travel required to keep $A_2$ constant during the scan, and is a measure of the topography $\Delta z_m(x,y)$.
For instance, it can be used to align images acquired in different fields.
The corresponding Fig.\,\ref{Fig:MFM_const_A2}(b) shows the first mode frequency shift $\Delta f_1(x,y,z_m(x,y))$.
It contains a pattern of spots congruent with  Fig.\,\ref{Fig:MFM_const_A2}(a), with an additional contrast pattern that is usually associated with the magnetic up and down domains (cf.~Ref~\cite{Hauet:2014fc}). 
The yellow/blue circles in Figs.\,\ref{Fig:MFM_const_A2}(b) and (c) indicate domes with an up/down magnetization. Magnetic contrast with high spatial resolution can also be obtained between the domes, but is generally difficult to assess whether such an area can switch its magnetization independently from that of the adjacent domes. In most cases an area between two domes changes its contrast from bright to dark if both adjacent domes switch from up to down. This could be caused by the magnetic exchange coupling of the film on the domes with the film at the location of the intermediate area, but could also be an artifact arising from a limited spatial resolution of the MFM. However, at least at a few positions -- highlighted in the blue insets in Figs.\,\ref{Fig:MFM_const_A2}(b), (c) and (d) -- the magnetization of the domes switch from up to down while part of areas between the two adjacent domes remains up (white) in a field of -153\,mT, but switch to down in a field of -406\,mT.

A salient feature of the above $\Delta f_1(x,y,z_m(x,y))$ images is the presence of dark spots at the centers of the domes, irrespective of the underlying magnetization orientation. 
In order to exclude that this contrast is of magnetic origin the area highlighted by the green frame in Fig.\,\ref{Fig:MFM_const_A2} was also measured in a field of +20\,mT. At such a field most of the film retains the down-state obtained at -406\,mT field, as known from prior work \cite{Hauet:2014fc}, but the magnetization of the tip has flipped. The latter can be seen from the dark contrast arising from the very few areas of the film with a magnetization direction changed from the down- to the up-state (see Fig.\,\ref{Fig:MFM_const_A2}(e)). The dark spots (domes) however remain the most prominent features in the $\Delta f_1$ image, although the tip magnetization is now antiparallel to the majority of the sample areas. If the contrast was of magnetic origin, the domes should now appear as white circles which is clearly not the case. Hence, apart from a small modification of the grey-level of the contrast at the location of the domes, the dark spots visible in all $\Delta f_1$ images are not of magnetic origin, but arise from a spatial variation of the van der Waals force, as already pointed our earlier work \cite{Hauet:2014fc}. 

However, in the present work $A_2$ was kept constant. One might thus expect that the local tip-sample distance, and thus also the van der Waals interaction remains constant. Then the domes should not be visible. The data taken in saturation (Fig.\,\ref{Fig:MFM_const_A2}(d)), however, shows that this is not the case. The reason for this discrepancy can be traced back to the different interaction length of two involved tip-sample forces: electrostatic ($U_{ac}\ne 0$) and attractive van der Waals ones, and thus to the respective interaction volumes.
The difference is confirmed by the departure from linearity of $\Delta f_1(A_2)$ (Fig.\,\ref{Fig:A2-df1}(b)), taking into account that the magnetic part of the interaction does not alter this fact.
Therefore, the van der Waals  contribution to $\Delta f_1$ will not remain constant when the tip traces lines of constant $A_2$, and the $z$-travel $\Delta z_m(x,y)$ required to keep $A_2$ constant will differ (slightly) from the true topography of the sample. 
Consequently, the domes will remain visible even if the image is acquired in a saturating field of -406\,mT  -- cf. Fig.\,\ref{Fig:MFM_const_A2}(d).

%

A more practical limitation of the {\em constant local height method} is that because of the small size of the $A_2$ the signal-to-noise ratio (SNR) available for $z$-control is limited.
As a consequence the $z$-position noise and thus also the $\Delta f_1$ noise in the image increases with the $z$-feedback speed, a fact that ultimately limits the sensitivity for small magnetic forces. Samples generating only weak stray fields are thus best measured at constant {\em average} height, i.e.~with slow proportional and integral parameters of the $z$-feedback.
They should be sufficiently fast to correct drift of the tip-sample distance but slow enough that localized topographical features encountered during the scan do not trigger a $z$-correction.
Note that appart from allowing to scan faster this method facilitates the quantitative interpretation of the MFM data \cite{Schendel00}.

\begin{figure}
\includegraphics{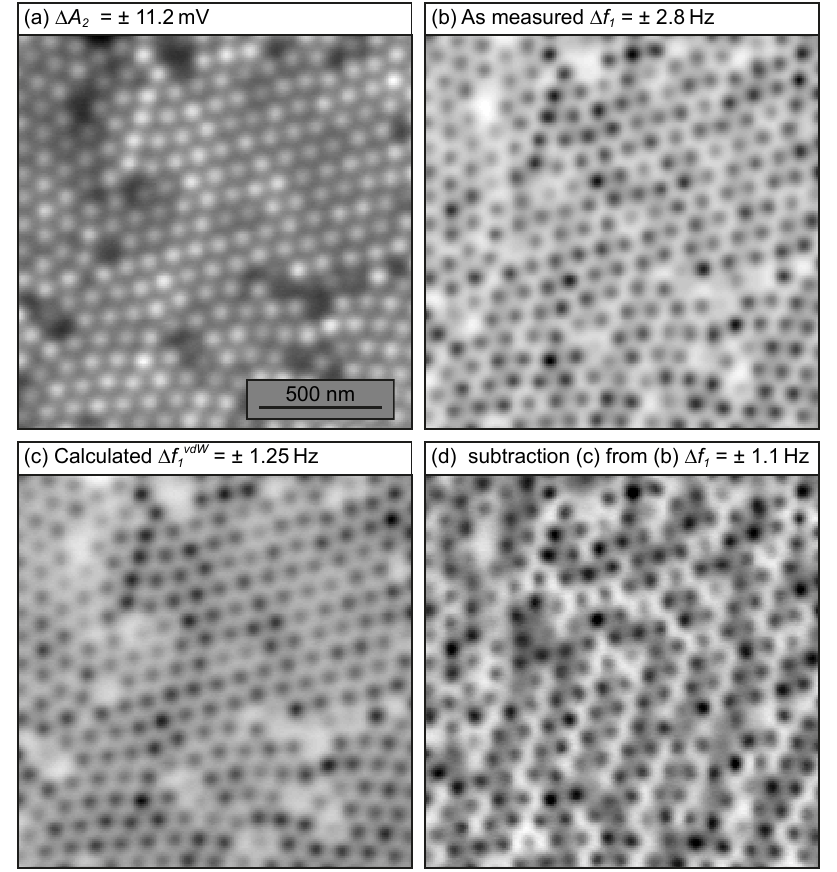}
\caption{\small
	(a) $A_2(x,y)$ data arising from topography-induced variations of the tip-sample distance.
	(b) Simultaneously measured $\Delta f_1(x,y)$ frequency shift data recorded at constant {\em average} tip-sample distance, i.e. with a slow $z$-feedback.
	(c) Van der Waals force induced variations of the frequency shift $\Delta f_1^{vdW}$ calculated from the $\Delta f_1(A_2)$-curve depicted in Fig.\,\ref{Fig:A2-df1}(c). 
	(d) Result of subtracting the data in (c) from that of (b), showing a magnetic interaction force-dominated image, which resembles the MFM image measured with a fast $z$-feedback shown in Fig.\,\ref{Fig:MFM_const_A2}(b).} 
\label{Fig:MFM_const_avg_z}
\end{figure}

Figure\,\ref{Fig:MFM_const_avg_z} displays one such {\em constant average height} measurement of the same area shown in Fig.\,\ref{Fig:MFM_const_A2}. The data were acquired in zero field, immediately after the constant $A_2$-scan in zero field was completed. The magnetic state of the sample thus is the same as shown in Fig.\,\ref{Fig:MFM_const_A2}(b).
The variations of $A_2$ (Fig.\,\ref{Fig:MFM_const_avg_z}(a)) using this measurement mode arise when local topography is (mostly) not compensated by the feedback.
These images of $A_2(x,y)$, taken at constant average height, can be translated into a frequency shift using the $\Delta f_1^{vdW}(A_2)$ dependence plotted in Fig.\,\ref{Fig:A2-df1}(b), which was derived from the measurements (Fig.\,\ref{Fig:A2-df1}(a)).
The result of this computation is shown in Fig.\,\ref{Fig:MFM_const_avg_z}(c).  
It can now be subtracted from the `raw' $\Delta f_1(x,y)$ data (Fig.\,\ref{Fig:MFM_const_avg_z}(b)) to enhance the relative weight of magnetic information in it.  
Figure\,\ref{Fig:MFM_const_avg_z}(d) displays the result, which qualitatively and quantitatively matches that acquired at constant local $A_2$ (Fig.\,\ref{Fig:MFM_const_A2}(b)).
It shows a substantial magnetic contrast with a weak background arising from the incomplete compensation of local van der Waals force variations.
Clearly, an ideally local treatment of van der Waals and electrostatic tip-sample interactions is an approximation that deteriorates when the sample topography is comparable to the tip that images it.
Future implementations of this techniques could rely on an explicit deconvolution, utilizing separate calibration measurements, to more perfectly compensate the topography-induced effects.
%

%

At this point it is convenient to note that the technique for distance control presented here could also prove useful for large area non-contact measurements of the Kelvin potential. The $z$-feedback that keeps $\Delta f_1$ (the frequency shift arising from van der Waals or interatomic forces) constant represents the topography only if the Kelvin potential is locally compensated. Usually this means that two feedbacks (the Kelvin- and the distance feedback) are arranged in series, rendering the selection of appropriate feedback parameters challenging and reducing the overall feedback speed. From equation Eq.\,\ref{eq:Eforce} follows that the $A_2$ signal is independent from the dc-potential (applied and Kelvin potential). Hence, a distance feedback using the $A_2$ signal will not be affected by the (slow) Kelvin feedback. 

More fundamentally, a distance feedback relying on van der Waals (or interatomic) forces requires the use of small tip-sample distances, an inherent difficulty in large area-scans of samples with substantial topography.

In conclusion, the capacitively controlled methods just discussed provide the experimenter with a robust technique for approaching, measuring and studying magnetic structures in the presence of non-negligible topography.
It can be seen that any Scanning Force Microscopy technique where $C(z)$ can be measured will benefit from the control of the tip-sample distance independently from non-capacitive tip-sample interaction forces of interest.

\acknowledgments
Support from the Swiss National Science Foundation, the CCMX, and Empa is hereby gratefully acknowledged. We thank L. Piraux, S. K. Srivastava, V. A. Antohe, M. Hehn, and T. Hauet for the preparation of the sample.


\end{document}